\documentclass[twocolumn, twocolappendix]{aastex701}
\usepackage{CJK} 
\usepackage{soul} 
\usepackage{amsmath}
\allowdisplaybreaks[4]
\usepackage[subject={Top1},author={\AA{}nsgar Lund},version=1]{pdfcomment}

\newcommand{\G}{\mathcal{G}}

\newcommand{\appropto}{\mathrel{\vcenter{\offinterlineskip\halign{\hfil$##$\cr\propto\cr\noalign{\kern2pt}\sim\cr\noalign{\kern-2pt}}}}}

\newcommand{\teff}{\ensuremath{T_{\rm eff}}}

\newcommand{\ar}{\ensuremath{a/R_{*}}}

\newcommand{\arpf}{\ensuremath{a_{\rm final}/R_{p}}}

\graphicspath{{./}{Figures/}}

\newcommand{\nmjbd}{31}

\newcommand{\njup}{88}

\newcommand{\nss}{39}

\newcommand{\nsnse}{12}

\newcommand{\nsysusedforanalysis}{170}
\newcommand{\nsysbinary}{92}

\newcommand{\jupcut}{117\pm9}
\newcommand{\sscut}{338\pm27}

\newcommand{\wsssig}{3.2}
\newcommand{\wjsig}{4.2}

\usepackage{url}
\usepackage{amsmath}
\usepackage{float}
\usepackage{rotating}
\usepackage{gensymb}
\usepackage{hyperref}
\usepackage{pgffor} 
\usepackage{soul}
\usepackage{changepage}
\usepackage{xfp} 

\begin{document}
\begin{CJK*}{UTF8}{gbsn}

\title{Warm Sub-Saturns Orbiting Single Stars Are Spin-Orbit Aligned}
\shortauthors{Wang et al.}

\author[0000-0002-0376-6365]{Xian-Yu Wang}
\altaffiliation{Sullivan Prize Postdoctoral Fellow}
\affiliation{Department of Astronomy, Indiana University, 727 East 3rd Street, Bloomington, IN 47405-7105, USA}
\email{xwa5@iu.edu}

\author[0000-0002-7846-6981]{Songhu Wang}
\affiliation{Department of Astronomy, Indiana University, 727 East 3rd Street, Bloomington, IN 47405-7105, USA}
\email{sw121@iu.edu}

 \correspondingauthor{Xian-Yu Wang}
 \email{xwa5@iu.edu}
 \correspondingauthor{Songhu Wang}
 \email{sw121@iu.edu}

\begin{abstract}

In this work, we show that warm sub-Saturns orbiting single stars are predominantly aligned, in contrast to hot sub-Saturns, which are frequently misaligned, with the two populations differing at the \wsssig$\sigma$ level. Because both populations are observed around cool stars, they are free from the ambiguity introduced by the $\teff-\lambda$ dependence. Together with the established alignment of warm Jupiters, this demonstrates, among single-star systems, that spin-orbit misalignment arises specifically in the close-in ``hot-Jupiter-analog'' regime, where tidal circularization is efficient ($\tau_e<\tau_{\rm age}$) and high-eccentricity migration is expected to operate.  We further find that the transition between aligned and misaligned sub-Saturns occurs at wider orbital separations ($\arpf \sim 340$) than for Jupiters ($\arpf \sim 120$), consistent with the expectation that the lower masses (smaller $M_p/M_*$) and stronger tidal dissipation (lower $Q_p$) of sub-Saturns allow them to be circularized into wider final orbits within their lifetimes. Taken together, these results provide the clearest direct evidence to date that, in single-star systems, spin-orbit misalignments are produced by high-eccentricity migration. If this framework is correct, spin-orbit misalignments may also emerge among hot-Jupiter analogs in other mass regimes, including hot brown dwarfs around hot stars at $\arpf \lesssim100$  and isolated hot super-Earths at $\arpf \lesssim1000$, with the corresponding transition locations shifted by the dependence of the orbital-circularization timescale on $M_p/M_*$ and  $Q_p$.

\end{abstract}

\section{Introduction} \label{sec:intro}

One of the strongest current lines of evidence that spin-orbit misalignments originate from high-eccentricity migration is that hot Jupiters are frequently misaligned \citep{Schlaufman2010, Winn2010, WinnFabrycky2015, Triaud2018, Albrecht2022}, whereas warm Jupiters around single stars are typically aligned \citep{Rice2022WJs_Aligned, WangX2024,Espinoza2025}. As recently pointed out by \citealt{Wu2023} and \citealt{Esposito2026}, this contrast is naturally expected if spin-orbit misalignments are generated by the same dynamical processes that excite orbital eccentricities \citep[e.g.,][]{Rasio1996, Weidenschilling1996, Wu2003, Wu2011}. Only planets driven to sufficiently small periastra experience tides strong enough to circularize within the system lifetime and become hot Jupiters, while retaining their spin-orbit misalignments, because orbital eccentricity and stellar obliquity are damped on different timescales \citep[e.g.,][]{Hut1981, Jackson2008, Adams2006, Albrecht2012, Lai2012, Li2016, zanazzi2024damping}. By contrast, the circularization timescale increases extremely steeply with periastron distance, so planets with slightly larger periastra cannot complete high-eccentricity migration and are not deposited as warm Jupiters through this channel efficiently \citep{Petrovich2016}. The observed warm-Jupiter population is therefore dominated by dynamically quieter formation pathways, such as disk migration \citep{Goldreich1980, Lin1996} or \textit{in situ} formation \citep{Boley2016, Batgyin2016}, that preserve primordial spin-orbit alignment.

This picture, however, is complicated by the strong dependence of stellar obliquity (true $\psi$ and sky-projected $\lambda$) on host-star effective temperature (\teff): hot Jupiters are misaligned almost exclusively around hot stars \citep{Schlaufman2010,Winn2010,Albrecht2012,Knudstrup2024}. Current Rossiter-McLaughlin \citep[RM,][]{Rossiter1924, McLaughlin1924} measurements of warm Jupiters, however, are drawn predominantly from cool-star systems, now defined as those with \teff\ cooler than the recently updated transition near 6500 K \citep[Kraft break;][]{kraft1967break, beyer_kraft_2024, Wang2026}. Consequently, while primordial spin-orbit misalignment has been investigated in a variety of contexts (e.g., compact multiplanet systems: \citealt{Albrecht2013} and \citealt{Morton2014}; star-disk systems: \citealt{watson2011}, \citealt{davies2019star}, and \citealt{Biddle2025}), current Rossiter-McLaughlin measurements of Jupiter-mass planets \textit{alone} do not yet distinguish whether the misalignments observed in hot Jupiters around hot stars reflect primordial hot-star misalignment or instead trace high-eccentricity migration, whose obliquity imprint may persist because tidal realignment is inefficient in hot stars \citep{Winn2010, Albrecht2012, Albrecht2022, Knudstrup2024, WangX2024}.

The stellar obliquity distribution of sub-Saturns helps clarify this picture. The discovery that sub-Saturns around cool stars can still be strongly misaligned (e.g., HAT-P-11, \citealt{Winn2010HATP11,Hirano2011HATP11,Sanchis2011HATP11,Deming2011HATP11, Bourrier2023}; HATS-38, \citealt{Espinoza2024}; GJ 436, \citealt{Bourrier2018,Bourrier2022}; GJ 3470, \citealt{Stefansson2022}; TOI-1842, \citealt{Hixenbaugh2023}; TOI-2374, \citealt{Yee2025TOI2374}; WASP-107, \citealt{Dai2017WASP107,Rubenzahl2021WASP107, Bourrier2023}; TOI-1135, \citealt{Dugan2025}) argues against the idea that misalignment is an outcome unique to hot stars. Moreover, recent work has shown that isolated sub-Saturn systems are more likely to be misaligned, whereas those in compact multi-planet systems are typically aligned \citep[e.g.,][]{Albrecht2013, Wang2018a, Zhou2018, WangX2022WASP148, Lubin2023, Zhang2025880c, Radzom2024, Radzom2025, Handley2026}, pointing to a dynamical rather than primordial origin for sub-Saturn stellar obliquities. However, while these clues favor a dynamical origin, they do not by themselves demonstrate that high-eccentricity migration is the responsible mechanism.

In this work, we show that warm sub-Saturns orbiting single stars are predominantly aligned, in contrast to hot sub-Saturns, which are frequently misaligned. Because both populations are observed around cool stars, they are free from the ambiguity introduced by the $\teff-\lambda$ dependence. Together with the established alignment of warm Jupiters, this demonstrates, among single-star systems, that spin-orbit misalignment arises specifically in the close-in ``hot-Jupiter-analog'' regime, where tidal circularization is efficient ($\tau_e<\tau_{\rm age}$) and high-eccentricity migration is expected to operate.  We further find that the transition between aligned and misaligned sub-Saturns occurs at wider orbital separations ($\arpf = \sscut$) than for Jupiters ($\arpf = \jupcut$), consistent with the expectation that the lower masses (smaller $M_p/M_*$) and stronger tidal dissipation (lower $Q_p$) of sub-Saturns allow them to be circularized into wider final orbits within their lifetimes \citep{Goldreich1966, Hut1981, Adams2006, Jackson2008}. Taken together, these results provide the clearest direct evidence to date that, in single-star systems, spin-orbit misalignments are produced by high-eccentricity migration.

\section{Sample Construction} \label{sec:SampleConstruction}

\begin{figure*}
        \centering
        \includegraphics[width=1\linewidth]{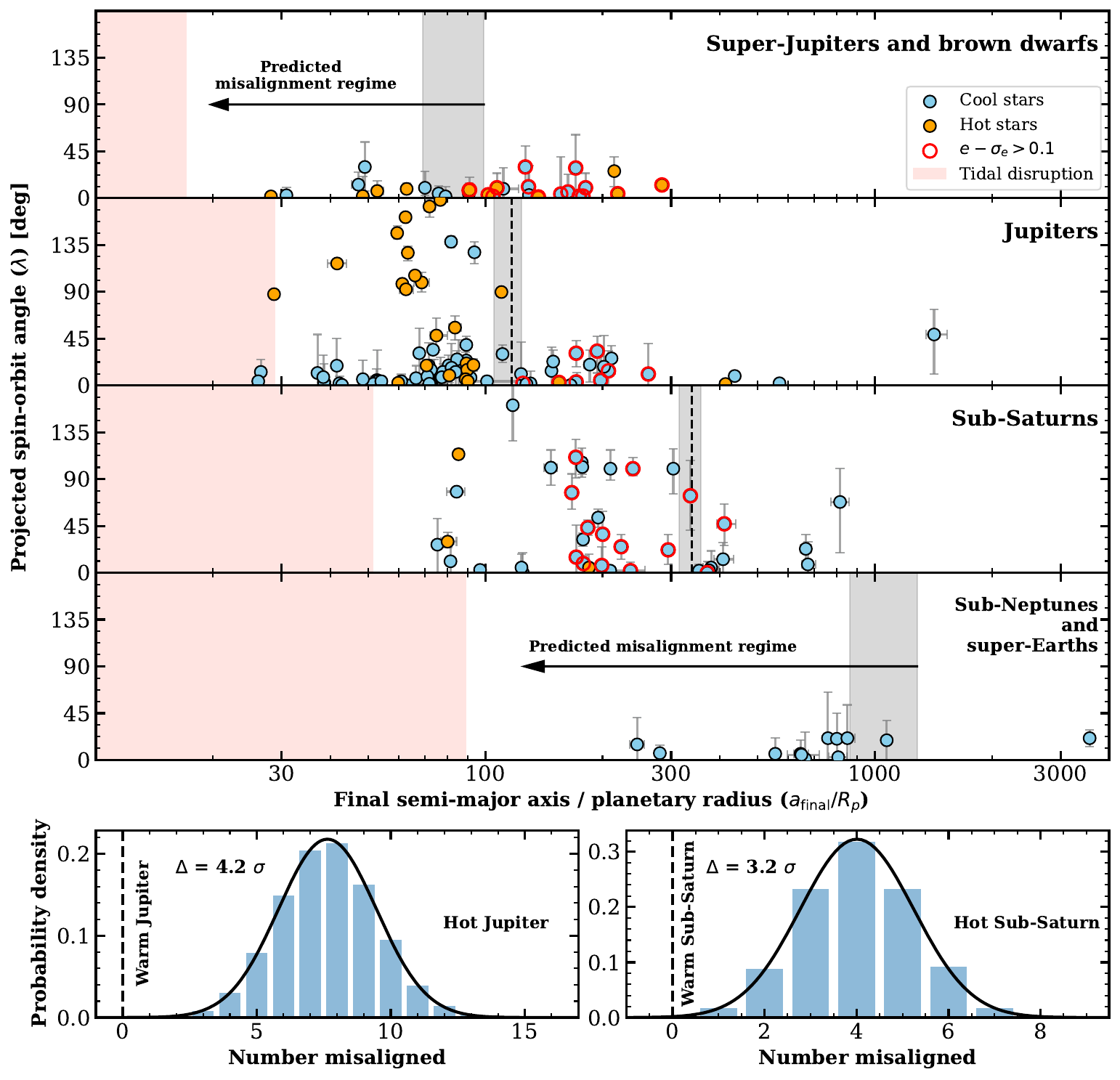}
\caption{The \textit{{upper}} four panels show the projected spin-orbit angle
$|\lambda|$ as a function of orbital separation scaled by the planetary radius
($\arpf$).
The sample is divided into four mass-ratio regimes from top to bottom:
super-Jupiters and brown dwarfs ($q > 0.002$), Jupiters
($0.0003 < q \leq 0.002$), sub-Saturns ($0.00005 < q \leq 0.0003$), and
sub-Neptunes and super-Earths ($q \leq 0.00005$).
Blue and orange symbols denote systems with host star effective temperatures
below and above 6500~K (at 1$\sigma$), respectively.
Red open circles indicate significantly eccentric orbits ($e - \sigma_e > 0.1$).
The tidal disruption zone is computed from the median mass ratio of each panel,
$a/R_p < 2.7\,(M_\star/M_p)^{1/3}$ \citep{Guillochon2011}.
Vertical dashed lines with shaded bands mark the $\arpf$ boundary that
optimally separates the $|\lambda|$ distributions of close-in and wide-orbit
systems, determined via a bootstrap Kolmogorov--Smirnov test.
In the first and fourth panels the gray bands are not measured but predicted
from the Jupiter boundary via the tidal-circularization scaling
($Q_p = 5\times10^{5}$ for super-Jupiters and brown dwarfs, $Q_p = 100$ for rocky
planets; see Section~\ref{sec:pred_cut}); misaligned orbits are expected
interior to them.
The \textit{{bottom}} two panels show the expected number of misaligned systems obtained by randomly drawing samples without replacement from the close-in population, with each draw matched in size to the observed wide-orbit population. The black dashed line marks the observed number of misaligned wide-orbit systems. The complete machine-readable system table underlying this figure can be found \href{https://github.com/wangxianyu7/Data_and_code/tree/main/WarmSubSaturnsTendToBeAligned}{here}.
}

        \label{fig:arplambda}
\end{figure*}

We constructed the sample using the system parameters from the Stellar Obliquity Catalog for Exoplanets and Brown Dwarfs (SOCat\footnote{Version dated July 17, 2026; \url{www.stellarobliquity.com}}), which was developed based on \citet{Wang2026WWB} and TEPcat\footnote{Version dated July 17, 2026; \url{https://www.astro.keele.ac.uk/jkt/tepcat/obliquity.html}} \citep{Southworth2011,Southworth2026}. To avoid observational biases introduced by different obliquity measurement techniques (see details in \citealt{Albrecht2022}), we restrict our analysis to measurements obtained from the Rossiter-McLaughlin effect, including the classical RM technique, Doppler Shadow/Tomography measurements \citep{Albrecht2007, Collier2010, Zhou2016, Johnson2017}, reloaded RM \citep[RRM,][]{Cegla2016ReloadedRM}, and RM Revolutions \citep[RMR,][]{Bourrier2021RMrevolutions} in order of priority.

Following \cite{Albrecht2022}, \cite{WangX2024}, and \cite{Wang2026}, we excluded measurements with large uncertainties ($|\sigma_{\lambda}| > 50 \degree$) and controversial cases: CoRoT-1 \citep{Bouchy2009CoRoT1,Pont2010}, CoRoT-19 \citep{Guenther2012}, HATS-14 \citep{Zhou2015}, HAT-P-27 \citep{Brown2012}, HD 3167  \citep{Dalal2019, Bourrier2021RMrevolutions,Teng2025}, WASP-1 \citep{Simpson2011a,Albrecht2011}, WASP-2 \citep{Triaud2010,Albrecht2011}, and WASP-49 \citep{Wyttenbach2017}.

Binary companions have been shown, both theoretically and observationally \citep[e.g.,][]{Wu2003, Fabrycky2007,Batygin2012, Naoz2012,Naoz2016, Hjorth2019, Espinoza2023, Rusznak2025, Wang2026}, to induce stellar obliquity across a wide range of planetary systems, further complicating the effort to determine the origin of spin-orbit misalignment. We therefore identified and excluded systems with known stellar companions using catalogs from \citet{Schwarz2016}, \citet{Fontanive2021}, and \citet{Badry2021}, as well as binary candidates, including HAT-P-50, which has a Gaia RUWE of 1.4, and K2-237, which lies in a crowded stellar field and is likely bound to Gaia DR3 6032746905476711680.  Although sufficiently wide companions may be dynamically decoupled from the present-day inner planetary system, \citet{Wang2026} surprisingly found that the stellar rotation and obliquity distributions of primary stars in \textit{known} binaries can differ from those of apparently single stars even at large binary separations.  Excluding systems with known stellar companions is therefore an empirically motivated sample-selection criterion, with the caveat that some apparently single stars may host undetected companions.

Moreover, our analysis is sensitive to both planetary and stellar masses, as the classification of exoplanet types in this work relies on the planet-star mass ratio; therefore, systems without accurate mass determinations (e.g., only upper limits are available) were excluded. 

This selection yields a single-star sample of \nsysusedforanalysis\, systems, after excluding \nsysbinary\, systems with stellar companions. We divide the sample into four mass-ratio regimes based on the planet-to-star mass ratio, $q = M_p/M_\star$, a parameter more directly linked to dynamical evolution than planetary mass alone \citep[e.g.,][]{Rusznak2025}: Sub-Neptunes and super-Earths ($q \leq 5 \times 10^{-5}$; \nsnse\, systems), sub-Saturns ($5 \times 10^{-5} < q \leq 3 \times 10^{-4}$; \nss\, systems), Jupiters ($3 \times 10^{-4} < q \leq 2 \times 10^{-3}$; \njup\, systems), and super-Jupiters and brown dwarfs ($q > 2 \times 10^{-3}$; \nmjbd\, systems). For a solar-mass host, these boundaries correspond approximately to 17\,$M_\oplus$, 93\,$M_\oplus$, and 2\,$M_{\rm J}$.

\section{Statistical Analysis}\label{sec:statistic}

\subsection{$\arpf$ as a Metric for Misalignment Origin}\label{sec:Defininghw}

The conventional distinction between `hot' and `warm' planets in stellar obliquity studies is typically based on the scaled orbital separation, $a/R_*$, with a boundary around $a/R_* \sim 8\text{--}10$ \citep[e.g.,][]{Albrecht2012, Rice2022WJs_Aligned, zanazzi2024damping, WangX2024}. This criterion is motivated by tidal realignment theory, in which the alignment timescale depends sensitively on orbital separation, scaling approximately as $(a/R_*)^{6}$ for cool stars with convective envelopes and $(a/R_*)^{17/2}$ for hot stars with radiative envelopes \citep{zahn1977reprint, Albrecht2012, Lai2012, Li2016}. In this framework, $\ar$ serves as a proxy for the efficiency of tidal realignment, and thus primarily traces the post-migration evolution of stellar obliquity.

However, for sub-Saturns, due to their lower planetary masses, post-migration stellar obliquities are only weakly affected by tidal realignment, as the tides they raise on their host stars are much weaker than those of Jovian planets. As a result, they are more likely to retain their primordial stellar obliquities and thus more directly trace their dynamical histories. If large stellar obliquities in single-star systems are predominantly generated by high-eccentricity migration \citep[e.g.,][]{Rasio1996, Weidenschilling1996, Wu2003, Fabrycky2007,Naoz2016, Wu2011,Petrovich2015, Rice2022,Wu2023,Kawai2025,Esposito2026}, then the key quantity becomes whether such migration can efficiently operate. High-eccentricity migration requires sufficient tidal dissipation to circularize initially eccentric orbits, with a circularization timescale $\tau_{\rm circ} \propto (\arpf)^5$, where $a_{\rm final} = a(1-e^2)$ is the final circularized semi-major axis assuming approximate conservation of orbital angular momentum \citep{Goldreich1966, Hut1981,Adams2006,Jackson2008}. In this sense, $\arpf$ serves as a proxy for the efficiency of high-eccentricity migration (see Equation~1 of \citealt{Adams2006}).

Therefore, we adopt $\arpf$ as the primary coordinate for empirically identifying the transition between hot and warm regimes in the observed stellar obliquity distribution. In this framework, systems at small $\arpf$ correspond to those in which high-eccentricity migration is efficient, while those at large $\arpf$ are less likely to have undergone such dynamically excited tidal migration.

\begin{figure*}
        \centering
        \includegraphics[width=1\linewidth]{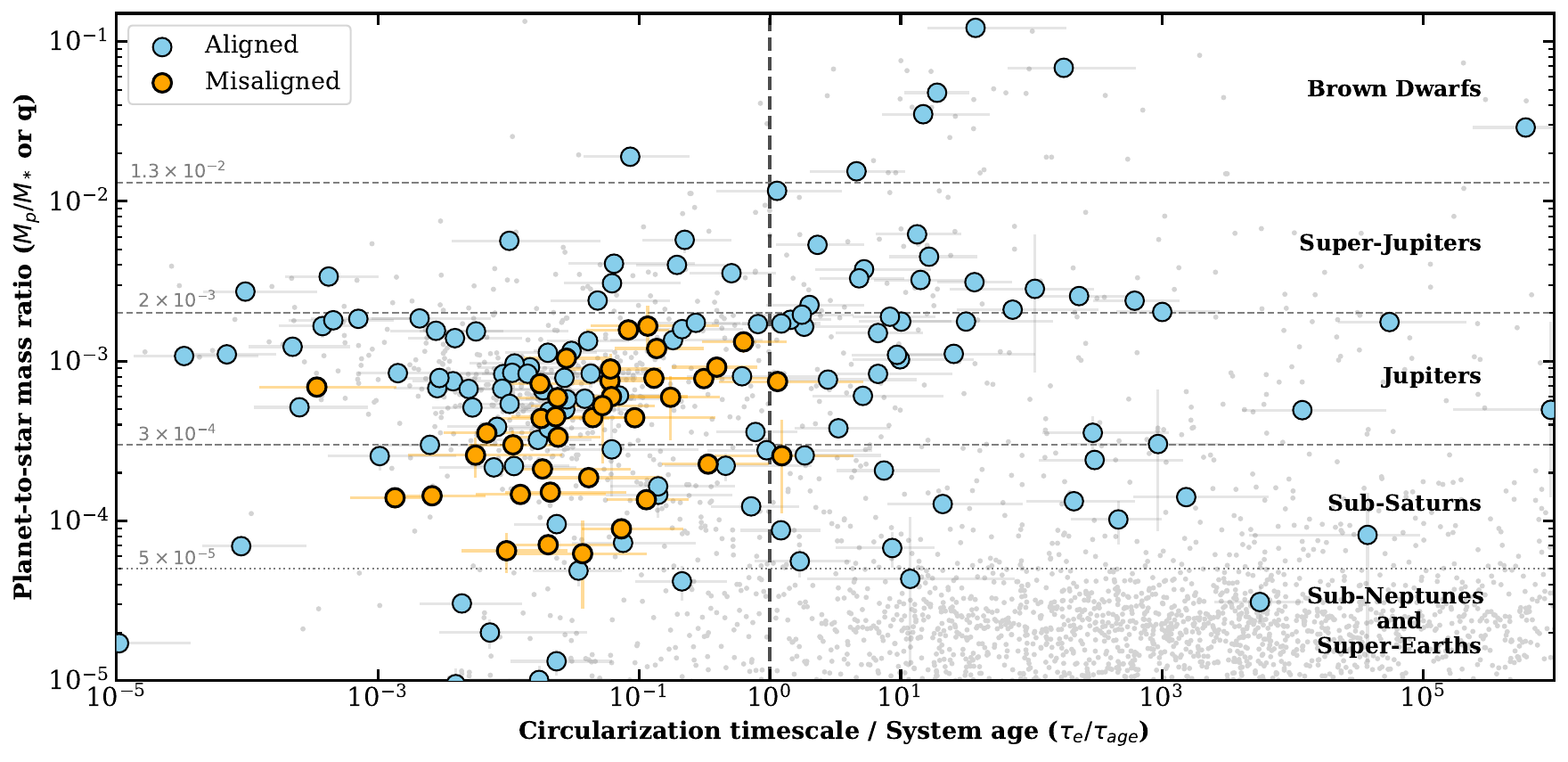}
        \vspace{-0.7cm}
\caption{
The planet-to-star mass ratio $q$ as a function of the eccentricity-damping timescale $\tau_e$, normalized by the system age ($\tau_{\rm age}$).
Blue and orange circles denote aligned and misaligned systems from the obliquity sample, respectively. Gray points show the broader exoplanet population from the PSCompPars table of NASA Exoplanet Archive
(\citealt{PSCompPars}, restricted to single stars with measured radii) and transiting brown dwarfs compiled from \cite{Barkaoui2025}.
The vertical black line marks $\tau_e = \tau_{\rm age}$. Horizontal dashed lines delineate the mass-ratio boundaries between
sub-Neptunes and super-Earths, sub-Saturns, Jupiters, super-Jupiters, and brown dwarfs. Systems to the left of the black line \textit{can} be produced through high-eccentricity migration. 
}\label{fig:ageq} 
\end{figure*}

\subsection{Derivation of Empirical Boundaries}\label{sec:boundary}

Figure~\ref{fig:arplambda} shows the $\arpf-\lambda$ distribution, illustrating a clear transition from misaligned hot planets to aligned warm planets in both the Jupiter and sub-Saturn populations. To determine the $\arpf$ boundary separating misaligned and aligned systems, we evaluated the $p$-values from two-sample Kolmogorov-Smirnov (K-S; \citealt{Hodges1958TheSP}) tests for a set of candidate cuts defined at the midpoints between adjacent unique values of $\arpf$.

For each candidate cut, we compared the $|\lambda|$ distributions of systems with $\arpf$ smaller than the cut and those with $\arpf$ larger than or equal to the cut using the two-sample K-S test as implemented in \texttt{SciPy} \citep{virtanen2020scipy}. We estimated the boundary location and its uncertainty through a bootstrap analysis. In each of 100,000 realizations, we resampled the systems with replacement and independently perturbed both $\arpf$ and $|\lambda|$ according to their reported measurement uncertainties. We then repeated the two-sample K-S analysis to determine the optimal cut. The resulting distribution of optimal cuts exhibits multiple peaks, reflecting the discrete and uneven sampling of systems in $\arpf$. We adopt the highest-$\arpf$ peak, which corresponds to the outer boundary of the observed misaligned population, and derive the boundary value and its uncertainty from this peak.

We applied the boundary derivation procedure to the Jupiter and sub-Saturn samples, obtaining $\arpf = \jupcut$ and $\arpf = \sscut$, respectively.

\subsection{Alignment Significance of Warm Sub-Saturns} \label{sec:sig}

To quantify the significance of the alignment tendency among warm sub-Saturns, we construct a reference distribution under the null hypothesis that warm sub-Saturns share the same obliquity distribution as hot sub-Saturns. We randomly draw eight sky-projected obliquity measurements without replacement from the hot sub-Saturn sample, equal to the total number of warm sub-Saturn systems with RM measurements, all of which are aligned. For each realization, we count the number of systems that satisfy the misalignment criterion ($|\lambda| - 3\sigma_{\lambda} > 0$ and $|\lambda| > 10\degree$; \citealt{Rice2022WJs_Aligned, WangX2024, Handley2026}). This procedure is repeated $100{,}000$ times to generate the expected distribution of misaligned systems under the null hypothesis. We approximate this distribution with a Gaussian characterized by mean $\mu$ = 4.01 and standard deviation $\sigma$ = 1.24. The observed number of misaligned warm sub-Saturn systems is zero, placing the sample $(\mu - 0)/\sigma \simeq$ \wsssig$\sigma$ below the expectation of the null distribution.

We perform the same analysis for the Jupiter sample and find that the alignment tendency of single-star warm Jupiters is significant at the \wjsig$\sigma$ level. Notably, although both Jupiters and sub-Saturns exhibit a separation-dependent obliquity transition, their host-star distributions differ: for Jupiters, almost all misaligned systems are found around hot stars, so the observed trend remains entangled with the $\teff-\lambda$ dependence, whereas for sub-Saturns, both aligned and misaligned systems are observed around cool stars, making the contrast free from this ambiguity. The corresponding distributions of the number of misaligned systems for both Jupiters and sub-Saturns are shown in the lower panels of Figure~\ref{fig:arplambda}.

\section{Discussion} \label{sec:Discussion}

\subsection{The Nature of $\arpf$--$\lambda$ Relation}
\label{sec:natureofarp}

As discussed in Section~\ref{sec:Defininghw}, $\arpf$ sets the tidal circularization timescale and therefore traces whether high-eccentricity migration can operate efficiently within the system lifetime. To quantitatively test the connection between spin-orbit misalignment and high-eccentricity migration, we compute $\tau_e$ for each system using the eccentricity damping formalism of \citet{Goldreich1966}, \citet{Hut1981}, and \citet{Jackson2008}:

\begin{equation}
\frac{1}{\tau_e} = \frac{1}{\tau_{e,p}} + \frac{1}{\tau_{e,\star}}
\end{equation}

\noindent The companion tide term is

\begin{align}
\tau_{e,p} = \frac{4 Q_{p}}{63}\left(\frac{a_{\rm final}^{3}}{G M_{*}}\right)^{1 / 2} \frac{M_{p}}{M_{*}}\left(\frac{a_{\rm final}}{R_{p}}\right)^{5}
\end{align}

\noindent while the stellar tide term is

\begin{equation}
\tau_{e,\star} = \frac{16 Q_\star}{171} \left(\frac{a_{\rm final}^3}{G M_\star}\right)^{1/2} \frac{M_\star}{M_p} \left(\frac{a_{\rm final}}{R_\star}\right)^5
\end{equation}

\noindent where $M_\star$, $R_\star$, $M_p$, $R_p$, and $a$ are the stellar mass, stellar radius, planetary mass, planetary radius, and orbital semi-major axis, respectively, and $Q_\star$ and $Q_p$ are the tidal quality factors of the star and the planet.

In this work, we assume a solar-like stellar tidal quality factor of $Q_\star = 4 \times 10^8$ for all host stars, following \citet{Penev2011} and \citet{Gallet2017}\footnote{\url{https://obswww.unige.ch/Recherche/evol/starevol/Galletetal17.php}}. For planetary and substellar companions, we adopt a nominal value of $Q_p = 5 \times 10^5$ for Jupiters, super-Jupiters, and brown dwarfs, $Q_p = 6 \times 10^4$ for sub-Saturns, and $Q_p = 100$ for sub-Neptunes and super-Earths \citep{Goldreich1966, Jackson2008, Kawai2025, Heller2010BDQ}. To propagate the substantial uncertainty in tidal dissipation into $\tau_e$, we assume a log-normal uncertainty of $0.3$~dex (a factor of $\sim2$) on $Q_p$ across all mass regimes, consistent with the empirical calibration $Q_{\rm Jup} = 4.9^{+3.5}_{-2.5} \times 10^5$ of \citet{Kawai2025}.

Figure~\ref{fig:ageq} shows the resulting $M_p/M_*$ - $\tau_e/\tau_{\rm age}$ - misalignment distribution. As \cite{Kawai2025} recently found, Jupiters with $\tau_e < \tau_{\rm age}$ can exhibit a wide range of stellar obliquities, while those with $\tau_e > \tau_{\rm age}$ tend to be aligned.  We find that sub-Saturn systems follow the same pattern. All misaligned systems in our sample have $\tau_e \leq \tau_{\rm age}$,
consistent with the interpretation that spin-orbit misalignment is
concentrated in the regime where tidal circularization is efficient and high-eccentricity migration is expected to operate.

Importantly, because both misaligned hot sub-Saturns and aligned warm sub-Saturns are observed around cool stars, the contrast between them is free from the ambiguity introduced by the $\teff-\lambda$ relation \citep{Schlaufman2010, Winn2010}. Together with the established pattern that hot Jupiters can be misaligned whereas warm Jupiters are generally aligned \citep[e.g.,][]{Wang2021, Rice2022WJs_Aligned, Wright2023, WangX2024, Espinoza2025}, this shows, among single-star systems, that spin-orbit misalignment arises specifically in the close-in ``hot-Jupiter-analog'' regime, where tidal circularization is efficient ($\tau_e<\tau_{\rm age}$) and high-eccentricity migration is expected to operate. The alignment of warm giants is most naturally explained as arising from a distinct evolution pathway rather than high-eccentricity migration. If this interpretation is correct, then the current alignment trend for warm Jupiters, although dominated by measurements around cool stars, should also extend to warm Jupiters around hot stars.

Moreover, we find that the transition between aligned and misaligned sub-Saturns occurs at larger scaled separations ($\arpf = \sscut$) than for Jupiters ($\arpf = \jupcut$), consistent with the expectation that the lower masses (smaller $M_p/M_*$) and stronger tidal dissipation (lower $Q_p$) of sub-Saturns allow tidal circularization to occur within the system lifetime out to larger $\arpf$, as suggested by the scaling $\tau_e \propto (\arpf)^5 (M_p/M_*) Q_p$ \citep{Goldreich1966,Hut1981,Adams2006}.

It has long been recognized, and variously explained \citep{Beust2012,Petrovich2020,Yu2024, Gao2025, Lu2025, IM2026}, that some misaligned sub-Saturn systems retain non-zero eccentricities despite having  $\tau_e < \tau_{\rm age}$ (famously, GJ~436; \citealt{Beust2012,Bourrier2018}). However, as recently pointed out by \cite{WangXYHomogeneousRM}, the deeper question is why misaligned hot Jupiters and misaligned hot sub-Saturns, both thought to have been delivered by high-eccentricity migration, show such different eccentricity outcomes, as highlighted by the red circles in Figure~\ref{fig:arplambda}: the former are almost always circularized, whereas the latter can remain \textit{eccentric} (see the compelling recent proposal by \citealt{Petrovich2026}, invoking thermally regulated viscoelastic tides in sub-Saturn interiors).

\subsection{Hidden Populations:\\ Misaligned Massive Companions and Small Planets}
\label{sec:pred_cut}

Across our sample, misaligned systems are concentrated in the hot-Jupiter and hot sub-Saturn regimes. In contrast, neither more massive companions (including super-Jupiters and brown dwarfs)\footnote{Although the $68\,M_{\rm J}$ brown dwarf HIP~33609~b, with $\lambda = 12.7 \pm 1.3^\circ$, formally satisfies our misalignment criterion, we do not regard a projected obliquity of only $\sim 10^\circ$ as compelling evidence that it is genuinely misaligned. We therefore classify the system as aligned in our analysis, consistent with the classification adopted by \citet{Vowell2026}.} nor small planets (including sub-Neptunes, super-Earths, and smaller planets) show clear evidence of spin-orbit misalignment around single stars in the analogous short-period regime. We argue that such systems may not yet have been uncovered observationally.

For massive companions, \citet{Rusznak2025} found that the alignment tendency holds regardless of orbital separation, and suggested that such objects may lack additional nearby companions capable of triggering dynamical instability \citep{Xu2024}. However, systems hosting multiple compact massive companions are rare but do exist. For example, TIC 279401253 \citep{Bozhilov2023} hosts two super-Jupiters with $P_b = 77\,\mathrm{d}, M_b = 6\,M_{\rm J}$ and $P_c = 155\,\mathrm{d}, M_c = 8\,M_{\rm J}$ orbiting a solar-like star ($M_\star = 1\,M_\odot$). Thus, tidal migration driven by dynamical instability in this regime cannot be fully excluded \citep{Wu2025}. Using the tidal quality factors described in Section~\ref{sec:natureofarp}, we therefore predict that super-Jupiters and brown dwarfs can likewise exhibit spin-orbit misalignments if their final scaled semi-major axes are sufficiently small ($\arpf \lesssim 69\text{--}99$, see Figure~\ref{fig:arplambda}). Because such massive companions can efficiently realign host stars with thick convective envelopes \citep{Albrecht2012, Lai2012}, this prediction is expected to apply only to hot stars, for which current RM measurements of close-in super-Jupiters and brown dwarfs remain extremely sparse (e.g., CoRoT-3, \citealt{Triaud2009}; GPX-1, \citealt{Giacalone2024}; HAT-P-34, \citealt{Albrecht2012}; HAT-P-69, \citealt{Zhou2019HATP69}; HATS-70, \citealt{Zhou2019HATS70}; HIP 33609, \citealt{Vowell2026}; KELT-1, \citealt{Siverd2012}; TOI-2109, \citealt{Wong2021}; TOI-3362, \citealt{Espinoza2023,Prinoth2024}; TOI-558, \citealt{Espinoza2025}; WASP-120, \citealt{Zak2024}). For companions in this mass regime, tides raised on the star may also contribute to orbital circularization. However, systems efficiently circularized by stellar tides may also undergo stellar-obliquity damping and become preferentially aligned; therefore, the prediction considers only tides raised on the companion.

For small planets, stellar obliquity measurements to date have been obtained almost exclusively for compact multi-planet systems \citep[e.g.,][]{Albrecht2013,WangX2022WASP148, Zhou2018, Dai2023,Radzom2024,Radzom2025,Handley2026}, which are expected to form and evolve quiescently within their natal disks. The apparent alignment of super-Earths may therefore primarily reflect an observational selection effect of this dynamically cool sample, rather than a genuine absence of misalignment. By analogy with the standard high-eccentricity migration picture for hot Jupiters, we suggest that if dynamical instability occurs in initially cold super-Earth chains, some planets may tidally migrate inward to become isolated, misaligned hot super-Earths. Adopting $Q_p = 100$, we find that such systems could be circularized into wider scaled separations, $\arpf \lesssim 1000$ (see Figure~\ref{fig:arplambda}). In contrast, sub-Neptunes may have tidal quality factors more similar to those of gas giants, implying that they can be delivered only to much closer-in orbits. This picture is relevant only for isolated planets, since compact multi-planet configurations largely exclude a high-eccentricity migration origin.

\section{Summary} \label{sec:Summary}

We find that warm sub-Saturns orbiting single stars are predominantly aligned, in contrast to hot sub-Saturns, which are frequently misaligned, with the two populations differing at the \wsssig$\sigma$ level. Because both populations are found around cool stars, they are free from the ambiguity introduced by the $\teff-\lambda$ dependence \citep{Schlaufman2010, Winn2010, Albrecht2012}. Together with a similar separation-dependent stellar obliquity transition that has previously been observed for Jupiter-mass planets \citep{Rice2022WJs_Aligned, WangX2024}, this demonstrates, among single-star systems, that spin-orbit misalignment arises specifically in the close-in ``hot-Jupiter-analog'' regime, where tidal circularization is efficient ($\tau_e<\tau_{\rm age}$) and high-eccentricity migration is expected to operate.

 We further find that the transition between aligned and misaligned sub-Saturns occurs at wider orbital separations ($\arpf = \sscut$) than for Jupiters ($\arpf = \jupcut$), consistent with the expectation that the lower masses (smaller $M_p/M_*$) and stronger tidal dissipation (lower $Q_p$) of sub-Saturns allow them to be circularized into wider final orbits within their lifetimes \citep{Goldreich1966, Hut1981, Adams2006}.

If this framework is correct, then spin-orbit misalignments should also emerge among hot-Jupiter analogs in other mass regimes, including hot brown dwarfs around hot stars at $\arpf \lesssim 100$ and isolated hot super-Earths at $\arpf \lesssim 1000$, with corresponding but shifted transition locations reflecting their different tidal properties. 

One remaining puzzle, recently highlighted by \citealt{Wang2026RMRefit}, is why misaligned hot Jupiters and misaligned hot sub-Saturns, both thought to have been delivered by high-eccentricity migration, show such different eccentricity outcomes: the former are almost always circularized, whereas the latter can remain eccentric \footnote{While this manuscript was under review, \citet{Petrovich2026} proposed a compelling possible explanation in which thermally regulated viscoelastic tides in sub-Saturn interiors sustain long-lived eccentric states.}.

\begin{acknowledgments}

We sincerely thank the reviewer for their constructive comments, which have helped improve the clarity and quality of the manuscript. This paper was inspired in part by a question raised by Eric Ford during the 2024 51 Pegasi b Summit: Why are warm sub-Saturns misaligned? We are grateful for helpful discussions with Cristobal Petrovich, Hareesh Gautham Bhaskar, Juan I. Espinoza-Retamal, Malena Rice, Gongjie Li, and Ji-Wei Xie. This work was supported in part by the NASA Exoplanets Research Program  NNH23ZDA001N-XRP (Grant No. 80NSSC24K0153), the NASA TESS General Investigator Program, Cycle~7, NNNH23ZDA001N-TESS (Grant No. 80NSSC25K7912), and the Heising-Simons Foundation (Grant \#2023-4050). We acknowledge funding support from grant JWST-GO-09025.010-A provided by NASA via the Space Telescope Science Institute under the JWST General Observers Program \#9025. Additionally, X.Y.W. acknowledges support from the Sullivan Prize Fellowship. This research was supported in part by Lilly Endowment, Inc., through its support for the Indiana University Pervasive Technology Institute.

\end{acknowledgments}

\facilities{NASA Exoplanet Archive}

\software{
\texttt{matplotlib} \citep{hunter2007matplotlib}, 
\texttt{numpy} \citep{oliphant2006guide, walt2011numpy, harris2020array}, 
\texttt{pandas} \citep{mckinney2010data}, 
\texttt{scipy} \citep{virtanen2020scipy}
}

\end{CJK*}
\bibliography{main}{}
\bibliographystyle{aasjournalv7}

\end{document}